\documentclass[11pt]{article}

\usepackage{cite}
\usepackage{graphicx}
\usepackage{amsmath,amssymb}
\usepackage[body={6in,8.5in}]{geometry}
\usepackage{color}
\usepackage{epstopdf}
\usepackage{subfigure}
\usepackage{url}
\usepackage{authblk}

\title{Mathematics and Climate Infographics: A Mechanism for Interdisciplinary Collaboration in the Classroom}

\author[1]{Ivan Sudakov \thanks{sudakov@math.utah.edu}}
\author[2]{Thomas Bellsky}
\author[3]{Svetlana Usenyuk}
\author[4]{Victoria Polyakova}
\affil[1]{Department of Mathematics, University of Utah}
\affil[2]{School of Mathematical and Statistical Science, Arizona State University}
\affil[3]{School of Art, Design and Architecture, Aalto University }
\affil[4]{Institute of Social and Political Sciences, Ural Federal University}
\bigskip
\date{}
\begin{document}
\maketitle
\begin{abstract}
Infographics are a form of data visualization combining data, information,  and statistics. Over the last ten years, infographics have become a popular method for displaying concise information, where infographics are a useful tool for classroom instruction. A high-quality infographic presents complex data in an aesthetically pleasing and simplistic format that allows student to understand more rapidly. 
Research within mathematics and climate science uses many elements of infographics. This work presents a series of electronic posters in an infographics style which explain hot topics in the mathematics of climate. These posters are designed to be used within standard undergraduate mathematical courses to provide students with concrete examples of how mathematics is incorporated within the climate sciences. 

\end{abstract}

\section{Infographics, Mathematics, and Climate}

Infographics are graphic visual representations of information, data, or knowledge intended to present complex information quickly and clearly  \cite {bib1}. They can improve cognition by utilizing graphics to enhance the human visual system's ability to see patterns and trends \cite {bib5}.  

Infographics are an eye-catching, engaging way to condense material into a more readily clear and accessible form. While many infographics focus on current events and technology trends, infographics are also becoming more common as an educational tool. There are many opportunities for incorporating educational infographics into the classroom, including using existing instructor-created visuals for instruction and as a form of assessment by having students create their own. 

Pedagogically, an interesting infographic can be used to introduce a new topic or to give a broad overview of a new subject. It can be used as a discussion starter, where students can answer questions or speculate about the given subject material. 
An infographic can also be used as a starting point for more in-depth research.

Climate science has a long history of visualizing information, as commonly seen with various maps of temperature and precipitation. In addition, modern applied mathematics is ubiquitous in the prediction and understanding of the Earth's climate. Visually combining such information is a natural tool for interdisciplinary education in mathematics and climate science. 

The work described in this manuscript combines expertise in mathematics and climate science to form a series of electronic posters in an infographics style which explain hot topics in the mathematics of climate for use in undergraduate mathematical courses. This series of infographics demonstrates applications of mathematics at the undergraduate level. These infographics are also useful at the high school level to peak interest in future undergraduate studies. In addition, these infographics are useful for more advanced students as a quick review of mathematical approaches used in climate science.

\section{Mathematics and climate infographics in the classroom: Innovative tools for instruction}

This project ``Mathematics and Climate Infographics" is devoted to offering classroom tools for teaching undergraduate topics in mathematics and statistics, with a focus on applications to climate. This effort is in support of the Mathematics of Planet Earth Year - 2013 (MPE2013) \cite {bib3}. A goal of MPE2013 is to involve high school and college students in mathematics and climate science through new educational materials developed for core mathematics courses. An additional aim of MPE2013 is that activities, materials, and didactic tools are freely available to any instructor in order to engage students in discovering the global impact of mathematics. In providing various materials to demonstrate applications of mathematics, an instructor also broadens the appeal of mathematics to students \cite {bib6}.  

Past experience shows the success of infographics in explaining modern mathematical problems. For example, during the World Mathematical Year 2000, a sequence of posters designed at the Isaac Newton Institute (INI) for Mathematical Sciences was displayed month by month in the trains of the London Underground \cite {bib4}. The main goal of the INI campaign was to bring mathematics to life, illustrating the wide applications of modern mathematics in all branches of science - physical, biological, technological, and financial.

Demonstration of mathematical applications is an important task in mathematical education. In particular, an undergraduate student may have a passing knowledge of topics in climate science; however, a sophisticated understanding often requires some level of knowledge of the underlying mathematical description of physical processes. In a typical undergraduate course, instructional time is often limited. An infographic can be useful in the classroom, as it is a quick way to transmit information.
 
The infographics from this project use color combinations and the arrangement of borders and space to provide a strong visual impact. Formulas and graphics, in combination with images of climatic objects, are used to convey a rapid understanding of various connections between mathematics and climate. The material of each infographic is presented in a succinct manner, so students will have a quick and understandable presentation of each topic. 

\begin{figure}[h!] 
\centering\includegraphics [scale=.3]{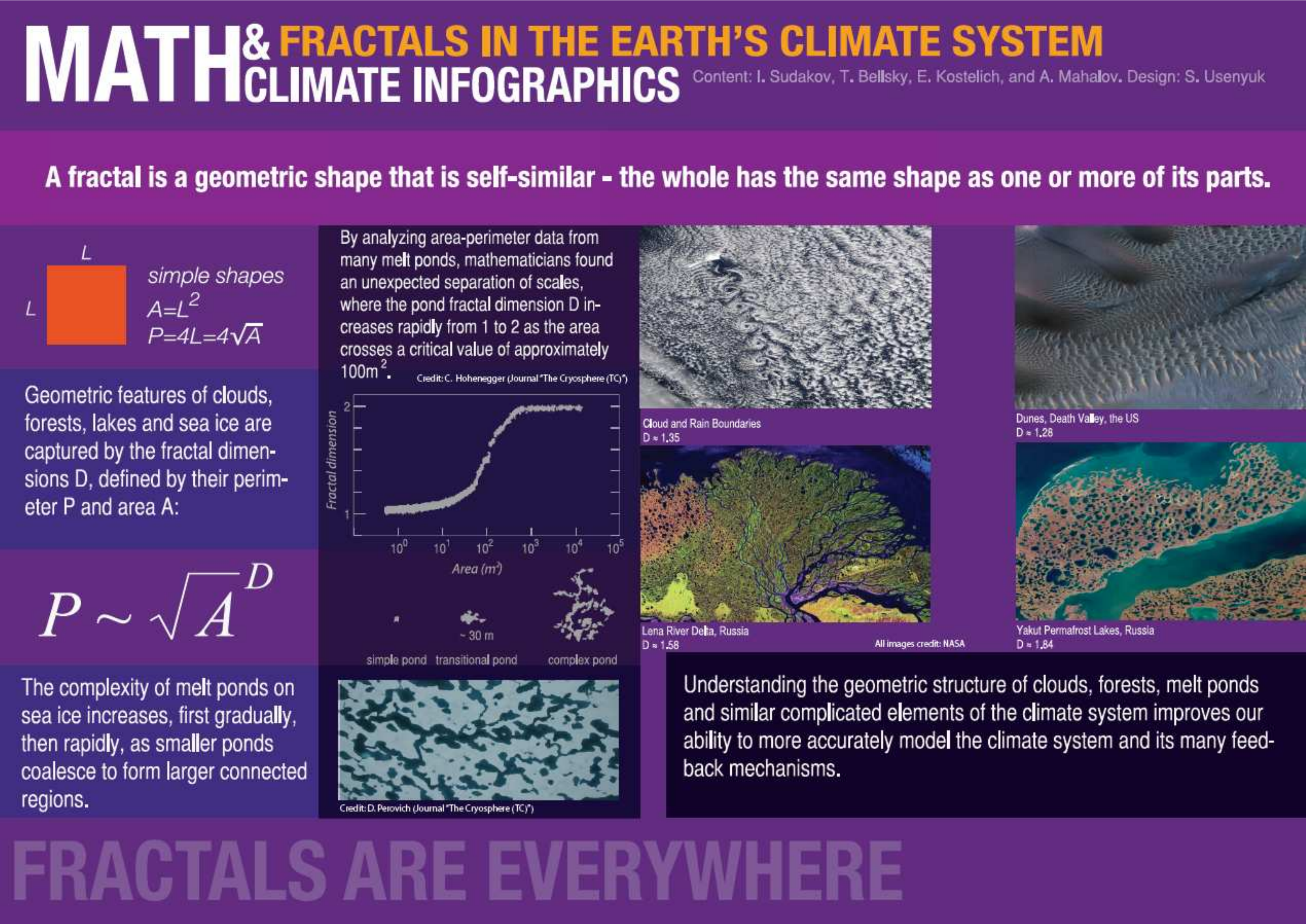}
\caption{\scriptsize{This math and climate infographic discusses fractals and their presence in climate phenomena. It presents a modern development of how fractal geometry is used in the important problem of Arctic climate change by studying the fractal dimension transition of melt ponds.}
\label{posterjul}}
\end{figure}

The infographic in Figure \ref{posterjul}
discusses the mathematical concept of fractals and how fractals relate to issues within the Earth's climate system. A fractal is any equation or pattern, that when visualized, produces a similar image when viewed at any spatial scale. This infographic describes how fractal-like patterns occur widely in nature and introduces the fractal dimension, a number which describes the fractal-like nature of an object. Specifically, it highlights the ubiquitous nature of fractals within many facets of the climate system. This infographic would be useful in a discussion of fractals in the undergraduate classroom, as it provides a real example of fractals in nature and exposes students to active climate research.
\begin{figure}[h!] 
\centering\includegraphics[scale=.3]{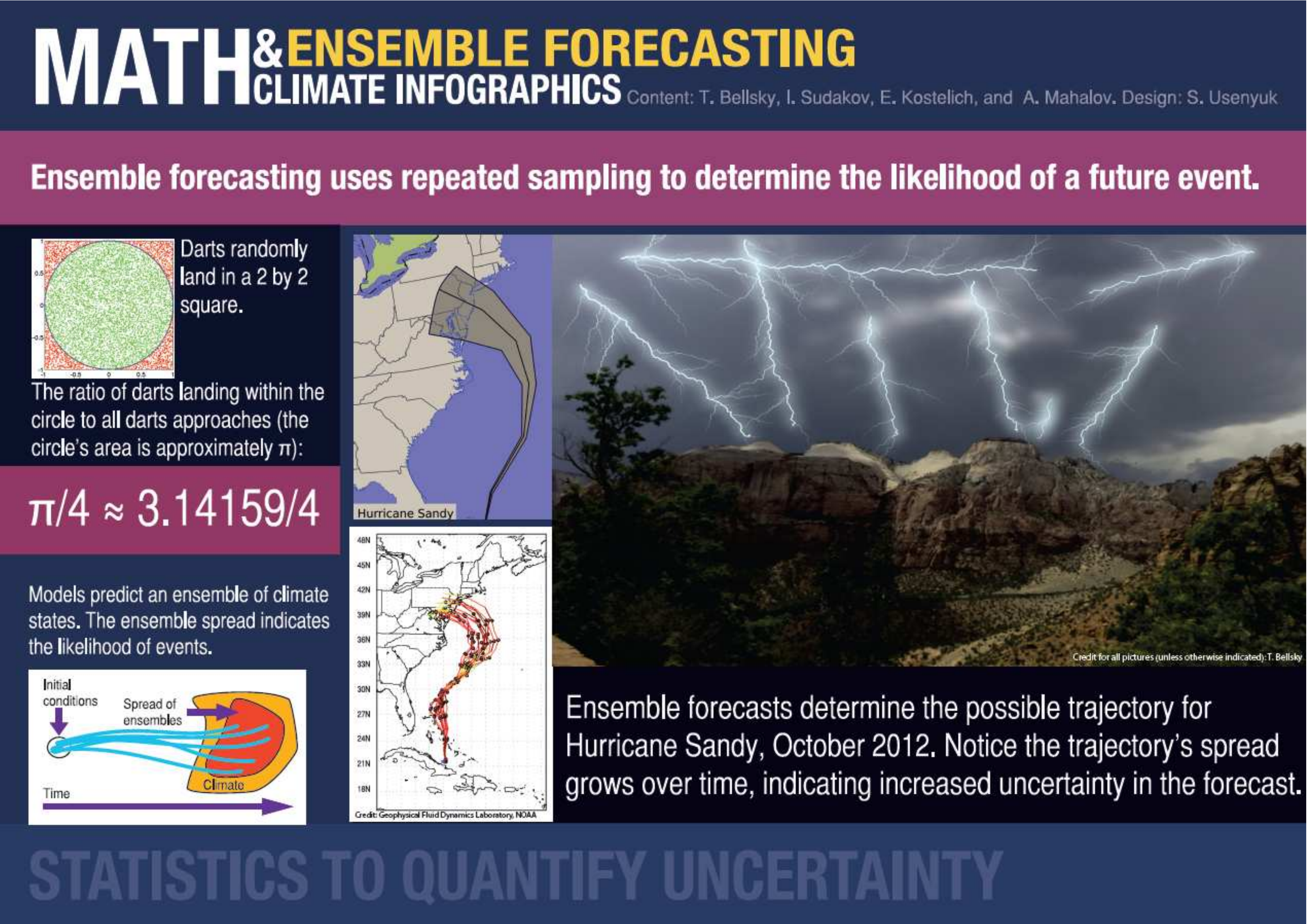}
\caption{\scriptsize{This math and climate poster discusses ensemble forecasting with applications to forecasting hurricanes.}
\label{posterjun}}
\end{figure}

The infographic in Figure \ref{posterjun} 
describes how statistical methods are used to forecast future events. Specifically, this infographic introduces the concept of ensemble forecasting, and how ensemble forecasting is used to determine the certainty of weather forecasts. This infographic then applies the introduced idea of ensemble forecasting to a very concrete example, where it examines the forecasted trajectory of Hurricane Sandy in October 2012 in the United States. 

\begin{figure}[h*] 
\centering\includegraphics[scale=.3]{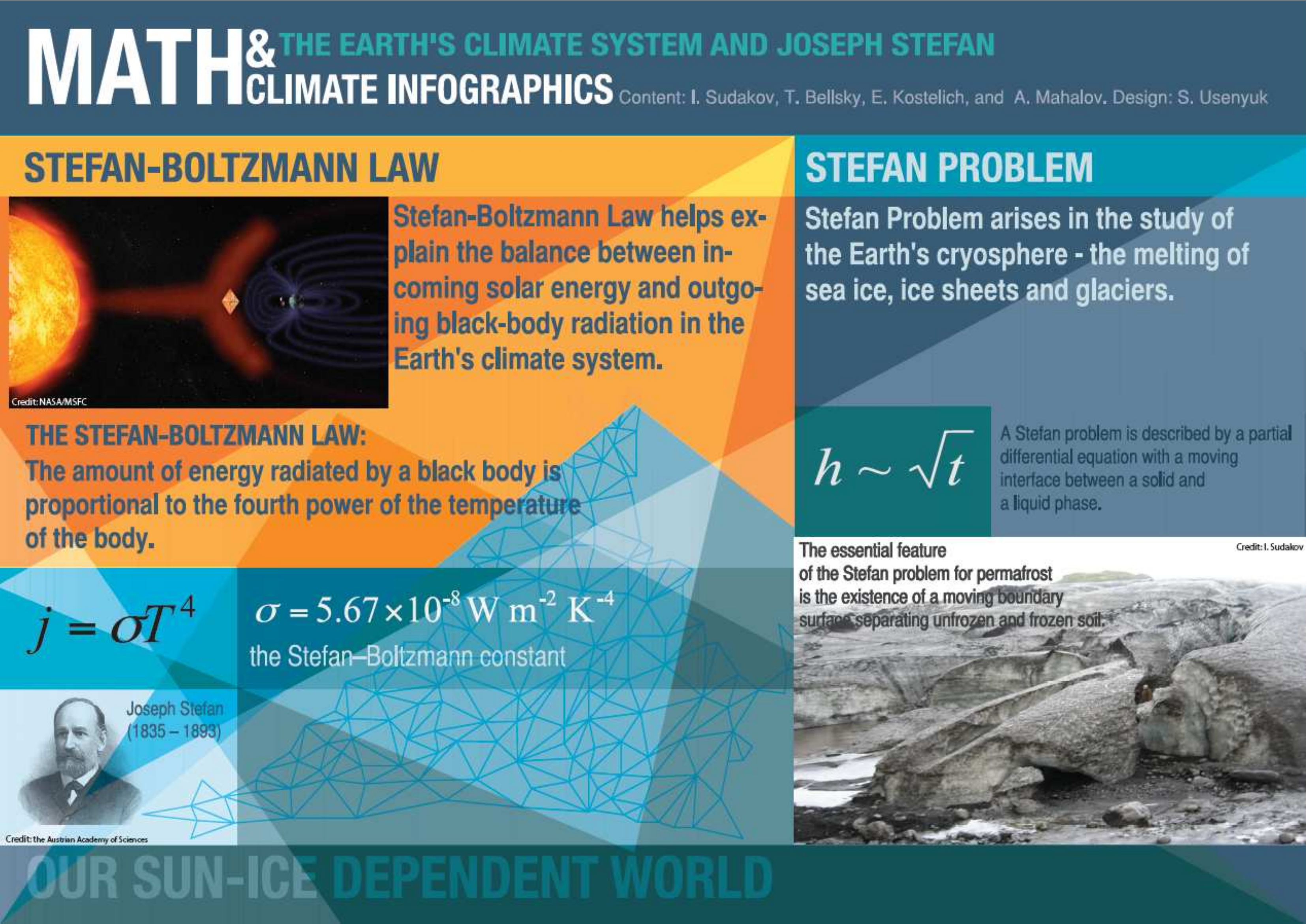}
\caption{\scriptsize{This math and climate infographic discusses Joseph Stefan's contribution to the study of ``warm" and ``cold" processes in the climate system.}
\label{postermar}}
\end{figure}

These infographics are also useful in demonstrating unexpected connections between mathematics and climate science. As an example, Joseph Stefan (Carinthian Slovene physicist and mathematician) made great contributions to the theoretical basis of Thermodynamics, but he was unaware that his research would be useful for modern climate change research. The infographic in Figure \ref{postermar}
describes both the Stefan-Boltzmann law, the basis of planetary energy balance equations, and the Stefan problem, which is used to study melting processes of ice. Initially, Stefan used this law to determine the temperature of the Sun's surface. Subsequently, this law has been applied to calculate planetary energy balance in conceptual climate models. The classical Stefan problem was developed to characterize the temperature distribution of a substance during a phase change. For example, Stefan's formula describes the dependence of ice-cover thickness in a particular region with this region's temperature record. This problem has subsequently been used to determine that the expected ice accumulation is proportional to the square root of the number of days when the air temperature is 
below freezing \cite{bib7}. The contribution of Stefan's work to modern science is a great example of interdisciplinary collaborations between physics, mathematics, and climate science. This particular infographic in Figure \ref{postermar} can be used to explain to students the importance, and often unexpected nature, of cross-disciplinary collaboration.  

This series of twelve infographic posters will be freely available to any educator. Based on our scientific interdisciplinary expertise, we have observed topics which are of current interest in modern climate system research, particularly: 

\begin{itemize}
  \item Ice Ages,
  \item Volcanic Winter, 
  \item The Earth's Climate System and Joseph Stefan,
  \item Urban Climate,
  \item Lorenz Butterfly,
  \item Ensemble Forecasting,
  \item Fractals in the Earth's Climate System,
  \item Climate Tipping Points,
  \item Carbon Dioxide Greenhouse Effect, 
  \item Climate Models, 
  \item Climate Networks, and 
  \item Climate Change and Other Planets.
\end{itemize}

\section{Survey and Discussion}
As part of this project, we have sampled undergraduate students about using infographics in the educational process. 
A total of $38$ undergraduate students were polled, all either mathematics majors or majors within the natural sciences.
In addition, the polled students represent two countries: the United States and the Russian Federation (37\% and 63\% respectively). Among the respondents, there were more females than males (65\% vs. 35\%).
 Also, a significant number of the polled students (87\%) were seniors who had already completed a substantial number of courses in mathematics and science. This survey shows interesting results about the quality of education, diversity in teaching methods, and new forms of visualizations in educational processes.

Surveyed students identified significant difficulties in learning within undergraduate courses as:
\begin{itemize}
  \item too much information: 53\%,
  \item too few illustrative examples: 35\%, and
  \item difficulty in perceiving information solely from lecture: 35\%.
\end{itemize}
All the selected difficulties above are associated with the need to improve properties like ``visualization" within classroom materials, where these difficulties listed above can be solved by using infographics. 

\begin{figure}[!htbp] 
\centering\includegraphics[scale=.7]{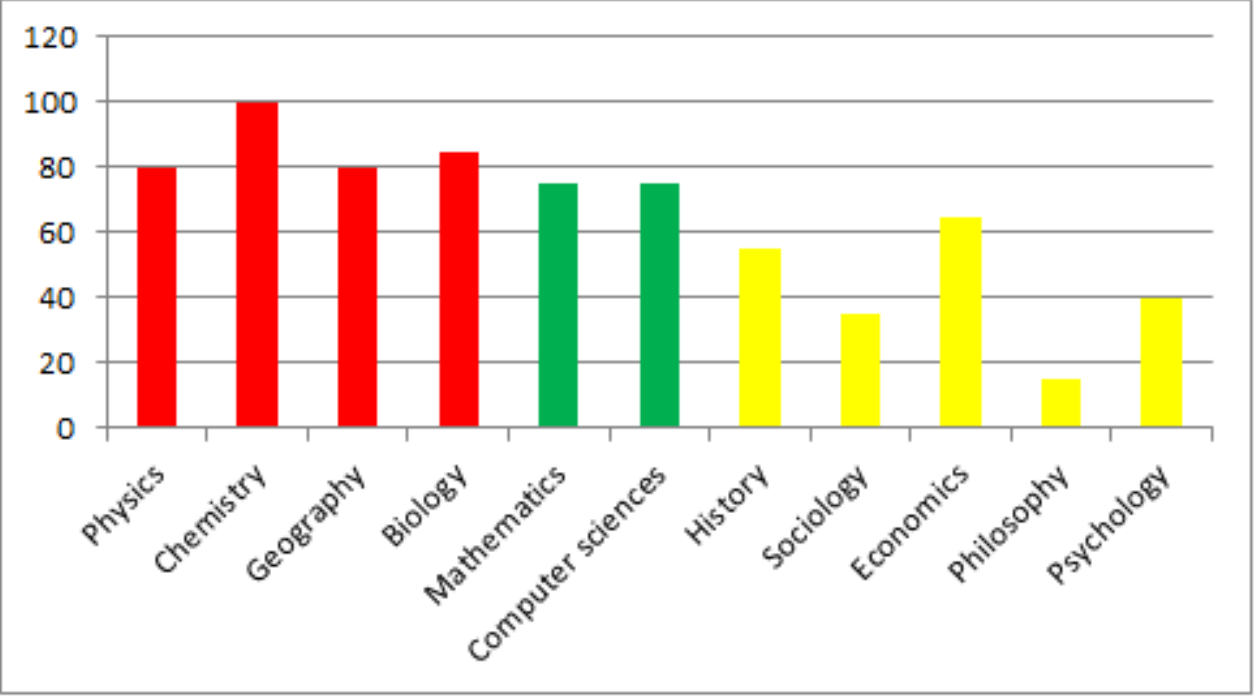}
\caption{\scriptsize{Disciplines for which infographics are most useful.}
\label{diagram}}
\end{figure}

According to polled students, the importance of infographics is highly dependent on a field of the study. As described in Figure \ref{diagram}, the surveyed students emphasized that the introduction of infographics in the educational process is extremely important for courses in the natural sciences such as chemistry, physics, geography, biology, general mathematics courses, and theoretical computer science. Additionally, surveyed students found infographics to be least useful in courses within the social sciences such as sociology, philosophy, and psychology.

More than 80\% of surveyed students have a positive attitude to the introduction of infographics in the educational process. A little less than half of the respondents (45\%) have previously encountered infographics in their undergraduate studies. In addition, 18\% of the surveyed students responded that infographics have been used in courses across many disciplines. Table  \ref{tab:imp} further presents surveyed students' responses to the importance of infographics in the educational process. 

\begin{table}[!htbp] 
\caption{\bf{Importance of using infographics, (\% of respondents)}}
\centering
\begin{tabular}{lr}
\hline
\bf{Importance} & \bf{\% of respondents}\\ \hline
Provide a visual representation
of material & 32.4\\ \hline
Contribute to a better perception 
of complex information & 26.5\\ \hline
Appropriate for the material of the studied course & 23.5\\ \hline
Concisely present a wealth of information & 14.7\\ \hline
Infographics are useful in all of the above & 2.9 \\ \hline
\end {tabular}
\label{tab:imp}
\end{table}
Additionally, respondents noted advantages of math and climate infographics as: 

\begin{itemize}
  \item Infographics are a brief presentation of a wealth of information - 65\%,
  \item Information is demonstrative and readable - 40\%, and
  \item Infographics show the relationship between different phenomena - 30\%.
\end{itemize}
Students emphasized the weaknesses of these infographics as: 

\begin{itemize}
  \item Information is not straightforward enough to understand - 17\% and
  \item Infographics are overloaded with details - 15\%.
  \end{itemize}
The criticism in the design of infographics may be caused by the amount of information that can be represented within a single poster.
Certainly, the amount of information contained in a poster is related to the complexity of the classroom material. Another possible cause of criticism of these infographics could be related to the instructor's qualifications. According to respondents, instructors who use infographics must have a certain set of competencies:
\begin{itemize}
  \item creativity - 70\%,
  \item a deep knowledge of the material in the discipline - 60\%,
  \item ability to analyze information - 55\%, and
  \item fluency in a variety of software programs - 35\%.
\end{itemize}
Additionally, from our own experience, we note that the instructor would certainty benefit if they are actively conducting interdisciplinary research. 
A final survey result found that half of students polled (50\%) believed infographics to be a desirable and effective tool in the educational process that would increase interest in learning within the mathematical sciences. In summary, these survey results find that infographics are useful in the classroom to demonstrate applications of mathematics. 

This collection of infographics promotes the incorporation of mathematics with modern climate science and education, and is useful in disseminating such important applications of mathematics to students. This kind of integration in turn conveys to the general public the importance of mathematical approaches to the understanding of climate processes.

\section*{Acknowledgments}
We are grateful for the financial support from the education and outreach mini-grant of the ``Mathematics and Climate Research Network" (MCRN), a network funded by the National Science Foundation linking researchers across the United States to develop mathematics to better understand the Earth's climate. In preparing this text, we have benefited from discussions with our colleagues: Tatyana Grechukhina,  Chris Jones, Hans G. Kaper, Eric Kostelich, Alex Mahalov, Monica Romeo, and  Mary Lou Zeeman. IS acknowledges the kind hospitality of the Isaac Newton Institute for Mathematical Sciences (Cambridge, UK) and of the Mathematics for the Fluid Earth 2013 programme, as well as the Dynasty Foundation and the Russian Federation President Grant (No. MK-128.2014.1) for their support. TB acknowledges support from the National Science Foundation under the grant DMS-0940314.

\end{document}